\pdfoutput=1
\documentclass[10pt, draftclsnofoot, onecolumn]{IEEEtran}
\usepackage{amsmath,amssymb}
\usepackage[dvips]{graphicx}
\usepackage{amsfonts}
\usepackage[mathscr]{eucal}
\usepackage{latexsym}
\usepackage{amsthm}
\usepackage{exscale}
\usepackage[mathscr]{eucal}
\usepackage{bm}
\usepackage[dvipsnames]{color}
\usepackage{cases}
\usepackage{color}
\usepackage{epsfig}
\usepackage{algorithm}
\usepackage{algorithmic}
\usepackage[verbose,nospace,sort]{cite}
\usepackage{tabularx}
\usepackage{multirow}
\usepackage{multicol}
\usepackage{balance}
\usepackage{caption}
\usepackage{subcaption}
\usepackage{setspace}
\usepackage{url}
\usepackage{cite}

\makeatletter
\renewcommand{\maketag@@@}[1]{\hbox{\m@th\normalsize\normalfont#1}}%
\makeatother
\scrollmode

\theoremstyle{definition}

\newtheorem{lemma}{Lemma}

\newtheorem{example}{Example}

\newcommand*{\QEDA}{\hfill\ensuremath{\blacksquare}}  

\usepackage{amsmath}

\begin{document}
\title{Impact of Channel Memory on the Data Freshness}
\author{Qixing~Guan and Xiaoli~Xu
\thanks{Q. Guan is with the School of Software Engineering, Southeast University, Nanjing 210096, China (e-mail: scout@seu.edu.cn).}
\thanks{X. Xu is with the School of Information Science and Engineering, Southeast University, Nanjing 210096, China (e-mail: xiaolixu@seu.edu.cn).}
}
\maketitle

\begin{abstract}
In this letter, we investigate the impact of channel memory on the average age of information (AoI) for networks with various packet arrival models under first-come-first-served (FCFS) and preemptive last-generated-first-served (pLGFS) policies over Gilbert-Elliott (GE) erasure channel. For networks with \textcolor{black}{Bernoulli} arrival model, we first derive the average AoI under the pLGFS queuing policy, and then characterize the AoI gap between FCFS and pLGFS policies. \textcolor{black}{For networks with \textcolor{black}{Bernoulli arrival and} generate-at-will arrival models, the AoI performances under the FCFS and pLGFS policies are derived explicitly.} For networks with periodic arrival model, we derive the closed-form expression for the average AoI under pLGFS \textcolor{black}{over a general GE channel} and propose a numerical algorithm for calculating that under FCFS efficiently. It is revealed that for pLGFS policy, the average AoI increases monotonically with channel memory $\eta$ at $\frac{\eta}{1-\eta}$ \textcolor{black}{over the symmetric GE channel}. For FCFS, the average AoI increases even faster \textcolor{black}{due to the queuing delay}, with an additional term related to the packet arrival rate.
\end{abstract}


\vspace{-0.15in}

\section{Introduction}
With the emerging of \textcolor{black}{the} time senstive applications, such as remote surgery and automatic driving, data freshness in communication systems has attracted a lot of research interest. Age of information (AoI) is one of the key metrics that describe the freshness attribution at the destination. The performances of the classic scheduling policies with the memoryless service time distribution have been investigated in \cite{kaul2012real, tripathi2019age, optLGFS, generalDis}. In \cite{kaul2012real}, the authors analyzed the average AoI of M/M/1, M/D/1 and D/M/1 queue models under the first-come-first-served (FCFS) discipline in the continuous system. The peak AoI for finite and infinite server queueing models under the preemptive last-generated-first-served (pLGFS) policy was considered in \cite{tripathi2019age}. The pLGFS policy was proved to be age-optimal with exponentially distributed service time\cite{optLGFS}. \textcolor{black}{Reference \cite{generalDis} and \cite{Kosta2021} derived general formulas for the stationary distribution of the AoI under various queuing policies in the continuous and discrete system, respectively.}

The work \cite{kaul2012real, tripathi2019age, optLGFS, generalDis, Kosta2021} mentioned above mainly focus on the channel which is time invariant, e.g., the erasure events in each slot are modelled by independent and identically distributed (i.i.d.) binary random variables. In practice, the communication channel usually has memory, or equivalently time correlation for consecutive packet transmissions. Gilbert-Elliott (GE) channel is a popular model adopted for channel with memory \cite{Gilbert}. Reference \cite{conti2Markov} derived the expression of the average AoI for the FCFS scheduling policy with Poisson arrival over GE channel in the continuous-time system by assuming the packet buffer size being zero or infinite. In \cite{timeCor}, the authors considered the  GE channel with periodic packet arrival in discrete system. With long term energy constrained, a threshold structure policy  was proposed to minimize the average AoI. The impact of channel memory on the end-to-end (E2E) latency of the communication network with GE channel has been investigated in \cite{Guan2112:Impact}. However, it is well known that optimizing the latency is fundamentally different from optimizing the AoI \cite{kaul2012real}, since the former focuses on the in-order delivery of all the packets, while the later emphasizes more on the update of the latest information.

To the best of our knowledge, the impact of channel memory under the classical packet arrival models and scheduling policies has not been explicitly derived or systematically studied. This letter aims to investigate the impact of channel memory on the data freshness by deriving the average AoI achieved under various packet arrival models and queuing policies. Specifically, we derive the closed-form expression for the average AoI achieved with pLGFS queuing policy for both Bernoulli and periodic arrival models, \textcolor{black}{along with FCFS and pLGFS for generate-at-will arrival model}. \textcolor{black}{We also propose} a novel approach for deriving the average AoI under FCFS by revealing the general AoI gap between pLGFS and FCFS policies, and design an efficient numerical method for calculating the average AoI under periodic arrival and FCFS policy.

The analytical results reveal that the average AoI achieved by pLGFS under the \textcolor{black}{symmetric} GE channel is larger than that in the memoryless channel with a constant gap $\frac{\eta}{1-\eta}$ for the Bernoulli, periodic \textcolor{black}{and generate-at-will} arrival, where $\eta\in[0,1)$ is the memory of the GE channel. The gap vanishes when the channel memory $\eta$ approaches $0$. For FCFS policy, the average AoI increases even faster with the channel memory, with an additional term related to the packet arrival rate. Under the same average arrival rate, the system with periodic packet arrival achieves better AoI than that with Bernoulli arrival \textcolor{black}{for the point-to-point link}.\\
\textcolor{black}{
The main contributions of this letter include:
\begin{itemize}
\item{We have analyzed the impact of channel memory on the AoI performance under various packet arrival models and queuing policies, which reveals that the AoI is more sensitive to the channel memory, as compared with the end-to-end latency \cite{Guan2112:Impact}.}
\item{	We have proposed a novel AoI analysis approach by characterizing the gap between the AoI evolution curves under pLGFS and FCFS policies. The approach can be generalized to other network models when the analysis for one of the queuing policies is much more challenging than the other one.}
\end{itemize}
}

\section{System Model}\label{sec:model}
We consider a slotted communication system where packets are generated at the beginning of each time slot. If it is transmitted immediately and not erased by the channel, it will be received at the end of the same slot. We consider the following packet arrival models:
\begin{itemize}
\item{\emph{Bernoulli arrival}, where a new packet arrives with probability $\lambda$ at each time slot, and no packet arrives with probability $(1-\lambda)$.}
\item{\emph{Periodic arrival}, where a new packet arrives every $K$ time slots, i.e.,  at the ${(iK+1)}$th slot, where $i=0,1,2,\cdots$, and no packet arrives at other slots. }
\textcolor{black}{\item{\emph{generate-at-will arrival}, where the transmitter can generate a packet whenever it wants \cite{updateOrWait}.}}
\end{itemize}

AoI is a metric reflecting the freshness of the information at the receiver, which is defined as the time interval between the current time slot and the generating slot of the \textcolor{black}{latest received packet}. We assume that the generating time slot is equivalent to arrival slot and the AoI is measured at the beginning of each time slot. Let $t_i$ denote the generating instant of the latest received packet at the receiver by the beginning of the time slot $t$. Then, the AoI of this time slot, denoted by $\Delta_t$, is given by
\begin{align}
\Delta_t = t - t_i
\end{align}
according to the definition \cite{kaul2012real}.

\textcolor{black}{AoI performance is closely related to the transmission policy. In this letter, two classical policies are considered:
\begin{itemize}
\item{\emph{pLGFS policy}, which is considered in a single packet queue. The latest arrival will be transmitted repeatedly until it is preempted by the next arrival or delivered to the receiver successfully.}
\item{\emph{FCFS policy}, which is considered with infinite buffer space at the transmitter. The packets are served according to their generating slots, which means that a packet can be transmitted only when all the previous packets have been successfully delivered.}
\end{itemize}}

To investigate the impact of channel memory on AoI, we consider the GE channel model, which is described by a two-state Markov chain. The Markov chain has a ``good" state and a ``bad" state, which is denoted as state $G$ and state $B$, with packet erasure probabilities $p_e^{G}$ and $p_e^{B}$, respectively. The GE channel transits from state $G$ to $B$ with probability $p$ and transits from state $B$ to $G$ with probability $r$, and its memory is defined as
\begin{align}
\eta=1-p-r. \label{eq:memory}
\end{align}
The GE channel has a persistent memory for $0<\eta<1$, and the average packet erasure probability is
\begin{align}
\textcolor{black}{\bar{p}_e}=\frac{r}{p+r}\textcolor{black}{p_e^G}+\frac{p}{p+r}\textcolor{black}{p_e^B}.
\end{align}
We assume $p_e^{G}=0$, $p_e^{B}=1$ in this letter and denote the channel by $GE(p,r)$ for convenience. \textcolor{black}{This model allows us to simplify the notations and focus on the impact of channel memory, which is only related with the state transition probability of the GE model.}

\section{The AoI with Bernoulli arrival Model}\label{sec:Ber}
In this section, we derive the average AoI with Bernoulli arrival model under the pLGFS and FCFS policies.
\subsection{Average AoI under pLGFS policy}\label{sec:BerLGFS}
For the pLGFS policy,  the AoI at the receiver is set to the age of the latest generated packet each time when the channel is in good state. Hence, the average AoI can be derived by analyzing the number of time slots between two good states and the age of the latest information at each good state. The result is summarized in Lemma~\ref{lemma2}.

\begin{lemma}\label{lemma2}
For a network with Bernoulli arrival packets at rate $\lambda$ over GE erasure channel $GE(p, r)$, the average AoI under pLGFS policy is given by
\begin{align}
{{\bar \Delta }_{BL}} = \frac{1}{\lambda } + \frac{p}{{r\left( {p + r} \right)}}.\label{eq:LGFS}
\end{align}
\end{lemma}
\indent \indent \emph{Proof:} Denote by $N$ the random variable representing the AoI at the receiver \textcolor{black}{at the moment when a packet is received. The age of the packet when it is received only depends on when it is generated. Under the pLGFS queuing policy, the old packets will always be preempted by the latest generated packets. Hence, if the packet is generated $n$ time slots before (with probability $\lambda$), it implies that it is not preempted by other packets during the subsequent $(n-1)$ time slots, which occurs with probability ${(1-\lambda)}^{n-1}$. Hence, the probability distribution of the AoI at the receiver when a packet is received is given by}
\begin{align}
P_N(n)  = \lambda{\left( {1 - \lambda } \right)^{n - 1}},n=1,2,\cdots
\end{align}

\textcolor{black}{Denote by $M$ the number of time slots between two consecutive good channel states, i.e., the number of time slots transiting from one good state (inclusive) to the next good state (exclusive). Hence, if the channel stays in good state with probability $(1-p)$, we have $m=1$. If $m\geq2$, it implies that the channel state transits to bad first and then to good again. Mathematically, we have:}
\begin{align}
{P_M}\left( m \right) = \left\{ {\begin{array}{*{20}{l}}
  {1 - p,}&{m = 1} \\
  {p{{\left( {1 - r} \right)}^{m - 2}}r,}&{m = 2,3, \cdots }
\end{array}} \right.\label{eq:GtoG}
\end{align}

\textcolor{black}{Consider the average AoI within the time slots between two consecutive packets delivered to the receiver. As illustrated in Fig.1 below, at the moment when a packet arrives, the AoI is equal to $n$ with probability $P_N (n)$. For the subsequent time slots, the receiver AoI $\Delta_r$ increases by $1$ at each time slot, until the next packet arrives. Hence, the sum AoI for the $m$ time slots within the segment is given by:
\begin{align}
\sum\limits_{n = 1}^\infty  {{P_N}\left( n \right)\left[ {n + \left( {n + 1} \right) +  \cdots  + \left( {n + m - 1} \right)} \right]}  = \sum\limits_{n = 1}^\infty  {{P_N}\left( n \right)\frac{{m\left( {2n + m - 1} \right)}}{2}}
\label{eq:segmentEq}
\end{align}
}

\begin{figure}[htb]
\centering
\includegraphics[scale=0.5]{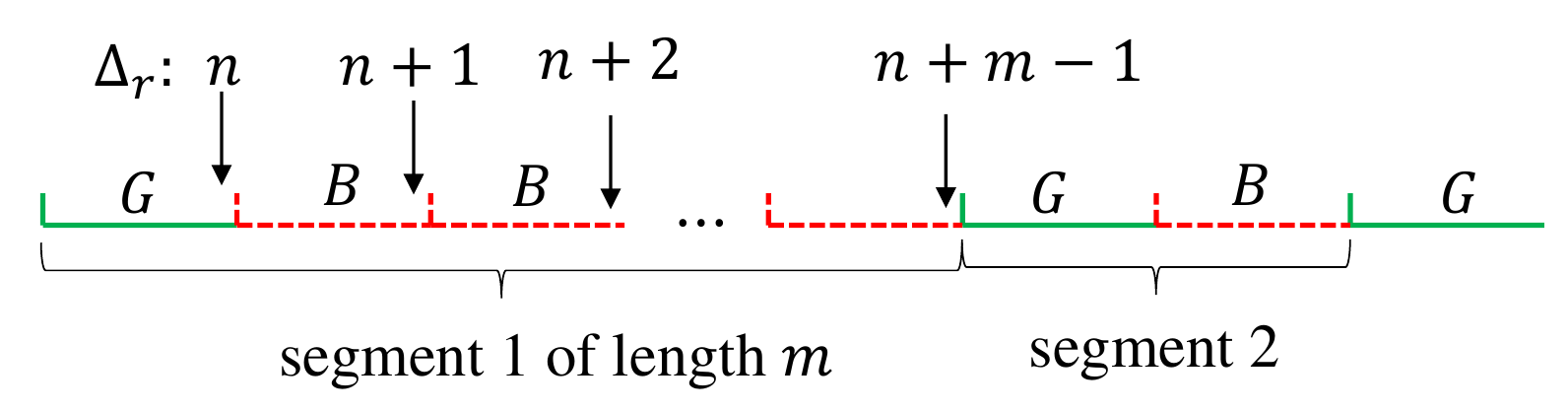}
\caption{An illustration of equation \eqref{eq:segmentEq}.}
\label{F:illustrateEq}
\end{figure}
\textcolor{black}{Since the whole transmission process can be divided into multiple segments, the sum AoI per segment can be computed as below:}

\begin{align}
\Delta_{BL-M}&=\sum\limits_{m = 1}^\infty  {P_M(m)\sum\limits_{n = 1}^\infty  {P_N(n)\frac{{m\left( {2n + m - 1} \right)}}{2}} }\notag\\
&= \frac{1}{\lambda }\left( {1 + \frac{p}{r}} \right) + \frac{p}{{{r^2}}}.\label{eq:deataM}
\end{align}
The expected number of time slots within the segment is:
\begin{align}
\mu _M = \sum\limits_{m = 1}^\infty  {mP_M(m)} = 1 + \frac{p}{r},\label{eq:muM}
\end{align}
and the average AoI can be derived as:
\begin{align}
{{\bar \Delta }_{BL}} = \frac{{{\Delta _{BL-M}}}}{{{\mu _M}}}.\label{eq:AoILGFS}
\end{align}
Substituting \eqref{eq:deataM}-\eqref{eq:muM} into \eqref{eq:AoILGFS}, we get the result in \eqref{eq:LGFS}.
\QEDA

When $p=r=0.5$, the channel $GE(p,r)$ is reduced to the random erasure channel with erasure probability $\bar{p}_e=0.5$, and \eqref{eq:LGFS} reduces to the Lemma 1 derived in \cite{xuAoIeffi}, i.e.,
\begin{align}
{{\bar \Delta }_{BL - 0}} = \frac{1}{\lambda } + 1.\label{eq:LGFS0}
\end{align}
To find the impact of channel memory, we consider the symmetric GE channel, i.e., ${p=r=\frac{{1 - \eta }}{2}}$, where $\eta$ is the channel memory defined in \eqref{eq:memory}. The average AoI in \eqref{eq:LGFS} reduces to a function of $\lambda$ and $\eta$ as
\begin{align}
{{\bar \Delta }_{BL - s}} = \frac{1}{\lambda } + 1 + {\frac{\eta }{{1 - \eta }}}.\label{eq:symLGFS}
\end{align}

Comparing \eqref{eq:LGFS0} and \eqref{eq:symLGFS}, we find that under the pLGFS policy, the channel memory brings the additional term in the AoI, i.e., $\frac{\eta}{1-\eta}$, which is monotonically increasing with $\eta$ \textcolor{black}{under the symmetric GE channel. For general asymmetric GE channel, given the channel memory, the receiver AoI depends on the transition probability, as given in the equation \eqref{eq:LGFS}.}

\subsection{Average AoI under FCFS policy}\label{sec3A}
\begin{figure}[htb]
\centering
\includegraphics[scale=0.5]{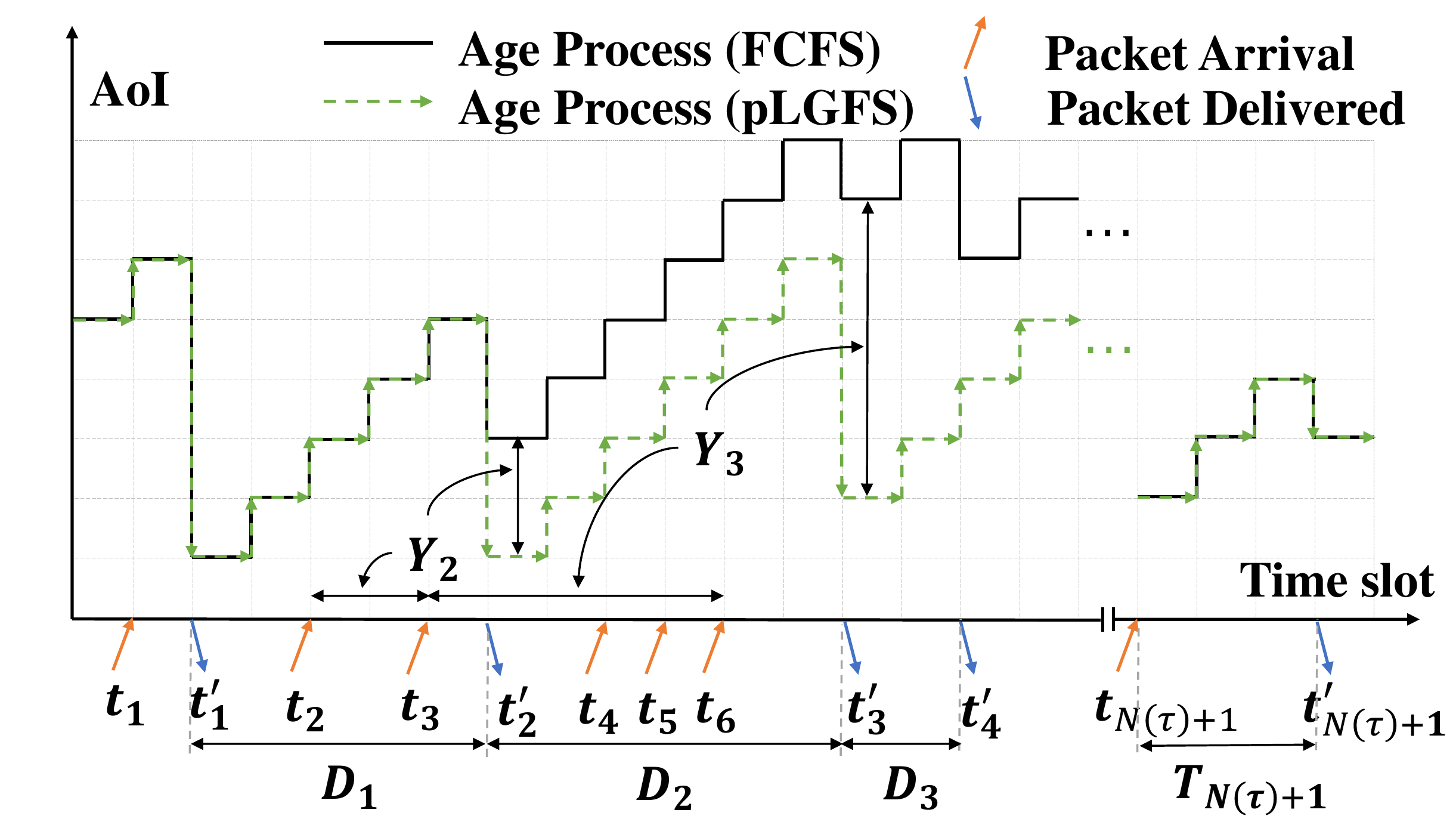}
\caption{An illustration example of the relationship between AoI curves under the FCFS and pLGFS policies. The $1$st packet is not preempted, while the $2$nd and $3$rd packet are preemped \textcolor{black}{by the packets that are generated} $2$ and $5$ time slots \textcolor{black}{later}, which lasts for $6$ and $2$ slots, respectively.}
\label{F:relation}
\end{figure}

Directly analyzing the AoI for FCFS policy under GE channel is rather challenging due to the coupling of the packet waiting time, service time with the queue status at the transmtiter. To this end, we propose a novel \textcolor{black}{analysis} approach by deriving the gap between the AoI curves for the pLGFS and FCFS policies. An illustration diagram for the AoI evolution under the pLGFS and FCFS policies is presented in Fig.~\ref{F:relation}.

\textcolor{black}{Assuming that the original transmission policy is FCFS, which is depicted in black solid line in Fig.~\ref{F:relation}.} Denote by $T_i={t_i^\prime}-t_i$ the E2E latency of the $i$th packet, where $t_i^\prime$ is the receiving time of the $i$th packet. $T_i$ is also referred to as ``system time" of the $i$th packet in this literature. Define \emph{inter-delivery time} ${{D_i} = t_{i + 1}^\prime - t_i^\prime, i \geq 1}$ as the time interval between two successive deliveries at the receiver, during which the AoI is valued as the elapsed time from generating slot $t_i$.

If the \textcolor{black}{original} policy is replaced by the pLGFS one \textcolor{black}{in green dashed line} and a new packet arrives during the system time of the $i$th packet, i.e., from $t_i$ to $t_i^\prime$, the old packet will be preempted in this case. Moreover, if the preemptive behaviour occurs repeatedly before $t_i^\prime$ (exclusive), the effective packet that contributes to the AoI curve is the last packet only, which is generated at ${t_w=\max \left\{ {t|t \in \left[ {{t_i},t_i^\prime} \right) \wedge {t_j} \ne \emptyset } \right\},j \ge i}$. The AoI curve under pLGFS policy is valued as the elapsed time from \textcolor{black}{$t_w$ accordingly}. 

Define \emph{preemption time} $Y_i= t_w- t_i,i \ge 1$ \textcolor{black}{the time difference between the generating slot of the preempted previous packet and that of the effective preempting current packet}. It is observed that the pLFGS policy introduces a constant AoI drop at value $Y_i\ge 1$, which lasts $D_i$ time slots. \textcolor{black}{As illustrated in Fig.~\ref{F:relation}, at the time instant $t_3^{\prime}$, the AoI of the pLGFS policy drops to $2$, i.e., the age of the $6$th arrival packets. However, the AoI of the FCFS policy only drops to $7$, i.e., the age of the $3$rd arrival packet. The AoI gap valued at $5$ contains the queuing delay of the $3$rd packet (from $t_3$ to $t_2^{\prime}$) under FCFS policy, which reflects the effect of queuing delay at the transmitter.} On the other hand, if no packet arrives during the system time, the preemptive event will not occur. The AoI curve under pLGFS still plots the time interval from $t_i$, which is identical with the FCFS case. The corresponding preemption time is $Y_i=0$ \textcolor{black}{and there exists no queuing delay}.

To summarize, \textcolor{black}{replacing the FCFS policy with pLGFS one under the same packet arrival model,} \emph{$Y_i=0$ means no preemption while ${Y_i\geq1}$ means a preemption during $T_i$, which contributes \textcolor{black}{the same} drop to the FCFS AoI curve, lasting $D_i$ time slots.} The accumulated AoI gap of the $i$th inter-delivery is
\begin{align}
\delta_i = D_i Y_i,
\end{align}
and the expectation could be derived by averaging the geometric parts between the two curves over time. 

Considering that there is no preemption before the first delivery $t_1^\prime$ or after the last one $t_{N\left( \tau  \right) + 1}^\prime$, the average AoI gap could be expressed as
\begin{align}
{\bar\Delta _{g} }=\mathop {\lim }\limits_{\tau  \to \infty }\frac{1}{\tau }{{\sum\limits_{i = 1}^{N(\tau )} \delta_i}=\lambda {\mathbb{E}\left[ {DY} \right]},}\label{eq:deltaGap}
\end{align}
where ${N\left( \tau  \right)}$ denotes the number of deliveries by time $\tau$. Hence, the average AoI under FCFS could be written as
\begin{align}
{\bar \Delta  _{F}} = {\bar \Delta _{L}} +  \lambda {\mathbb{E}\left[ {DY} \right]},\label{eq:gapExpression}
\end{align}
where ${\bar \Delta _{L}}$ is the average AoI under pLGFS.

We then focus on the preemption time $Y$ and inter-delivery time $D$ in \eqref{eq:gapExpression}. For  $Y_i\geq1$, the packet being served under FCFS will be preempted under pLGFS. In this case, the random variable $D$ only depends on the channel parameter, which is equal to the time it takes to transit from one good state to the next good state for the first time. It means that $D$ is independent on $Y$ when $Y_i\geq1$. With the results in \eqref{eq:GtoG} and \eqref{eq:muM}, we have
\begin{align}
{\mathbb{E}\left[ {D|Y\geq1} \right]}=1+\frac{p}{r}.\label{eq:EDGivenYgeq1}
\end{align}
Considering the fact that $Y_i=0$ does not contribute to the expectation, we write
\begin{align}
\mathbb{E}\left[D_{i} Y_{i}\right] =\mathbb{E}\left[D_{i} \mid Y_{i} \geq 1\right] \mathbb{E}\left[Y_{i}\right].\label{eq:DandY}
\end{align}

Given a typical packet transmission process, whose system time is ${T_i=t}$. The meaningful value of $Y_i=y$ ranges from $1$ to $(t-1)$, resulting in
\begin{align}
\mathbb{E}\left[Y_{i}\right]=\sum_{t=2}^{\infty}\left(\sum_{y=1}^{t-1} y P_{Y \mid T}(y \mid t)\right) P_{T}(t).\label{eq:EYexp}
\end{align}
The probability mass function (PMF) of the system time $T$ is given by
\begin{align}
P_{T}(t)=\left\{
\begin{aligned}
&\frac{r-(p+r) \lambda}{(p+r)(1-\lambda)}, &t=1 \\ 
&\frac{p}{p+r} \frac{r-(p+r) \lambda}{(1-\lambda)^{2}}\left(\frac{p+(1-p-r)(1-\lambda)}{1-\lambda}\right)^{t-2}, &t\ge2
\end{aligned}
\right.
\label{eq:Tn}.
\end{align}
\indent \indent \emph{Proof:} Please refer to Appendix~\ref{appendixA}.\QEDA

For a given system time ${t}$, ${Y_i=y}$ means a preemption occurs with $y$ slots after its generating slot, which requires no packet arrival after the preemption. We derive the conditional PMF of $Y$ given $T$, i.e.,
\begin{align}
{P_{Y|T}}\left( {y|t} \right) = \left\{ {\begin{array}{*{20}{l}}
{{{\left( {1 - \lambda } \right)}^{t - 1}},}&{y = 0}\\
{\lambda {{\left( {1 - \lambda } \right)}^{t - 1 - y}},}&{y = 1,2, \cdots, t - 1}
\end{array}}. \right.\label{eq:YdisGivenT}
\end{align}

With the results of \eqref{eq:LGFS}, \eqref{eq:EDGivenYgeq1} - \eqref{eq:YdisGivenT} substituted into \eqref{eq:gapExpression}, we derive the average AoI of Bernoulli arrival packets under the FCFS policy, as presented in Lemma \ref{lemma1}.

\begin{lemma} \label{lemma1}
For a network with Bernoulli packet arrival model at rate $\lambda$ over GE erasure channel $GE(p, r)$, the average AoI under FCFS queue policy is given by
\begin{align}
\bar{\Delta}_{BF}=\frac{1}{\lambda } + \frac{p}{r}\left[ {\frac{1}{{p + r}} + \frac{{{\lambda ^2}}}{{\left( {r - \left( {p + r} \right)\lambda } \right)\left( {r - \left( {p + r} \right)\lambda  + \lambda } \right)}}} \right].\label{eq:FCFSeq}
\end{align}
\end{lemma}

When $p=r=0.5$, the channel $GE(p,r)$ reduces to the random erasure channel with i.i.d. erasure probability ${\bar{p}_e=0.5}$, and \eqref{eq:FCFSeq} is reduced to the Lemma 1 derived in \cite{xuAoIeffi}, i.e.,
\begin{align}
{\bar \Delta _{BF-0}} = \frac{1}{\lambda } + 1 + \frac{{4{\lambda ^2}}}{{1 - 2\lambda }}.
\end{align}

Specially, the average AoI achieved by the FCFS policy over symmetric GE channel with memory $\eta$ is given by
\begin{align}
{{\bar \Delta }_{BF-s}} = \frac{1}{\lambda } + 1 +\underbrace{\frac{\eta}{{1 - \eta }} }_{\bar\Delta_1(\eta)}+ \frac{{4{\lambda ^2}}}{{ {1 - 2\lambda }}}\underbrace{\frac{1}{{\left( {1 - \eta } \right){{\left( {1 - \left( {1 - 2\lambda } \right)\eta } \right)}}}}}_{\bar\Delta_2(\eta)}.\label{eq:FCFSsys}
\end{align}
It is observed that when $\eta\in[0,1)$, ${\bar \Delta }_{BF - s}$ converges if and only if $\lambda  \in \left( {0,0.5} \right)$. This is consistent with expectation since the AoI will grow to infinity if the packet arrival rate exceeds the channel capacity, which is $0.5$ here.  For a given parameter $\lambda$, the additional AoI term includes two parts, satisfying $\bar\Delta_1(\eta) \ge 0$ and $\bar\Delta_2(\eta)\ge 1$. The former is the same as that under pLGFS policy, while the latter is the additional term related to the packet arrival rate $\lambda$ \textcolor{black}{and caused by the queuing delay}. For networks operate close to the capacity, we have $(1-2\lambda)\rightarrow 0$. The additional AoI penalty, $\bar\Delta_2(\eta)$, is proportional to the factor $\frac{1}{1-\eta}$, which grows faster as compared with that under pLGFS.

\section{The AoI with generate-at-will arrival Model}\label{generate-at-will}
\textcolor{black}{
In this section, we consider the transmission over the networks with generate-at-will arrival model, where the transmitter can generate a packet whenever it wants. Under the pLGFS policy, the transmitter generates a packet at each time slot to catch up with the next good state, whose AoI performance is equivalent to that under Bernoulli arrival with $\lambda=1$, i.e.,
\begin{align}
{\Delta _{wiL}} = 1+\frac{p}{r(p+r)}.\label{eqx:generWillpLGFS}
\end{align}
The corresponding result under symmetric GE channel is
\begin{align}
{\Delta _{wiL-s}} = 2+\frac{\eta}{1-\eta}.\label{eqx:generWillpLGFSsym}
\end{align}
Under the FCFS policy, to avoid the possible queuing delay, the packet is generated upon its previous packet is delivered to the receiver. With the relative results derived in section \ref{sec:BerLGFS}, the difference here is that the PMF of initial receiver AoI value $N$.\footnote{The notations in this subsection share the corresponding meanings with those in section \ref{sec:BerLGFS}.} With generate-at-will arrival model, $N$ measures the time interval from the previous good state, which means that the random variable $N$ shares the same PMF with $M$, i.e., $P_N(n)=P_M(n)$, leading to
\begin{align}
{\Delta _{wiF - M}} = {\left( {1 + \frac{p}{r}} \right)^2} + \frac{p}{{{r^2}}}.\label{eqx:generWill}
\end{align}
Substituting \eqref{eqx:generWill} into RHS of \eqref{eq:AoILGFS}, we get the result as Lemma~\ref{lemma4}.
\begin{lemma}\label{lemma4}
For a network with generate-at-will arrival packets over GE erasure channel $GE(p, r)$, the average AoI under the FCFS policy is given by
\begin{align}
{{\bar \Delta }_{wiF}} = 1+\frac{p}{r} + \frac{p}{{r\left( {p + r} \right)}}.\label{eq:ageOfWill}
\end{align}
\end{lemma}
For symmetric GE channel where $p=r=\frac{1-\eta}{2}$, the result is
\begin{align}
{{\bar \Delta }_{wiF-s}} = 3+\frac{\eta}{1-\eta},\label{eq:gen_at-will}
\end{align}
which insreases with the channel memory $\eta$. We note that the additional AoI under the two policies are both $\frac{\eta}{1-\eta}$. This is because there is no queuing delay when the packets are generated at will. 
}

\section{The AoI with Periodic Arrival Model}\label{sec:Deter}
In this section, we consider the transmission over the networks with periodic arrival model, where the packets are generated every $K$ time slots.
\subsection{Average AoI under pLGFS policy}\label{sec:perLGFS}
The average AoI of periodic arrival packets under the pLGFS policy is presented in Lemma \ref{lemma3}.
\begin{lemma}\label{lemma3}
For a network with periodic arrival parameter $K$ over GE erasure channel $GE(p, r)$, the average AoI under pLGFS policy is given by
\begin{align}
{{\bar \Delta }_{pL}} = \frac{{K + 1}}{2} + \frac{p}{{r\left( {p + r} \right)}}.\label{eq:pLGFSorigin}
\end{align}
\end{lemma}
\indent \indent \emph{Proof:} \textcolor{black}{The derivation of the average AoI under periodic arrival is similar to that with the Bernoulli arrival, except that the packet is generated} at the transmitter every $K$ slots. \textcolor{black}{Hence, the distribution of the age when the packet is delivered is given by}
\begin{align}
P_N ( {n} ) = \frac{1}{K},n = 1,2,\cdots,K.
\end{align}
\textcolor{black}{The expected accumulated age for the time slots between two consecutive packet arrivals is hence given by}
\begin{align}
\Delta_{pL-M}= \frac{1}{2}\left( {1 + \frac{p}{r}} \right)\left( {K + 1} \right) + \frac{p}{{{r^2}}}.\label{eq:deataM2}
\end{align}
Substituting \eqref{eq:deataM2} into RHS of \eqref{eq:AoILGFS}, we get the result in \eqref{eq:pLGFSorigin}.\QEDA

When ${p=r=0.5}$, i.e., $GE(p,r)$ reduces to random erasure channel with erasure probability $\bar{p}_e=0.5$, \eqref{eq:pLGFSorigin} reduces to the result of P-LCFS D/M/1/0 model in \cite{generalDis}, but in discrete correspondence, i.e.,
\begin{align}
{{\bar \Delta }_{pL-0}} = \frac{{K + 1}}{2} +1\label{eq:pLGFS0}.
\end{align}

The average AoI achieved by the pLGFS policy over symmetric GE channel with memory $\eta$ is given by
\begin{align}
{{\bar \Delta }_{pL-s}} = \frac{{K + 1}}{2} +1+ {\frac{\eta }{{1 - \eta }}}.\label{eq:pLGFSsys}
\end{align}
We note that the additional AoI due to channel memory for periodic arrival is the same as that for Bernoulli arrival. Besides, for packet arrival process with the same average rate, i.e., $\lambda=\frac{1}{K}$, periodic model should achieve smaller AoI than Bernoulli model by a factor of $\frac{K-1}{2}$.

\subsection{Average AoI under FCFS policy}
The relationship between the achievable AoI under FCFS and pLGFS policy still works for periodic arrival model. However, the PMF of preemption time depends on that of system time, which is intractable for the periodic packet arrival model. Hence, we derive the AoI by its relation with the E2E latency in this case, which is solved numerically  in Algorithm~\ref{alg1}.

Reference \cite{kaul2012real} has proved that with periodic inter-arrival time $K$, the average AoI under FCFS policy has a constant gap with the E2E latency, for any distribution of the service time. Mathematically, we have
\begin{align}
\bar{\Delta}_{pF}= \mathbb{E}\left[ T \right] + \frac{K-1}{2}.\label{AoIperK}
\end{align}

Hence, deriving the average AoI is equivalent to deriving the expected E2E latency, \textcolor{black}{with queueing delay included}. Denote by $t_{n,G}$ and $t_{n,B}$ the expected E2E latency experienced by a packet which arrives when the transmitter has queue length $n$, and the channel is in good and bad state, respectively. According to the analysis in our previous work \cite{Guan2112:Impact},  $t_{n,G}$ and $t_{n,B}$ can be explicitly expressed as
\begin{align}
&t_{n,G}=1 + n\left( { 1  + \frac{p}{r} } \right),\nonumber \\
&t_{n,B}=1 + \frac{1}{r}  + n\left( {1  + \frac{p}{r} } \right).
\end{align}

Further denote by $p_{n,G}$ and $p_{n,B}$ the probabilities for the queue size at the transmitter being $n$, with the channel in good and bad state, respectively. The expected E2E latency can be expressed as
\begin{align}
\mathbb{E}\left[ T \right] = \sum\limits_{n = 0}^\infty  {\left(t_{n,G}{p_{n,G}} + t_{n,B}{p_{n,B}}\right)}\label{perE2Elatency},
\end{align}
The key for deriving $\mathbb{E}[T]$ is to derive the probability distributions $\{p_{n,G},p_{n,B},n=0,1,\cdots\}$, \textcolor{black}{which could be calculated numerically.} 

\textcolor{black}{Without loss of generality, we consider a segment of time from the $1$st to the $(K+1)$th slot, where the time indexes are relative}. Denote by $P(K,n,s^{\prime}|s)$ the probability that \textcolor{black}{there are $n$ good channel states during $K$ consecutive time slots},\footnote{The value of $P(K,n,s^{\prime}|s)$ can be calculated in a recursive manner by following the formulas given in \cite{GEmodelRecur}, which is omitted here for brevity.} \textcolor{black}{leaving} the channel state at the \textcolor{black}{$(K+1)$th} time slot being $s'$ given the $1$st being $s$, where ${s,s'\in\{G,B\}}$. For a typical packet, assuming that it arrives \textcolor{black}{when the queue length is $i$,} the probability that the next packet arrives \textcolor{black}{at the $(K+1)$th slot with queue length being $n$} is equivalent to the probability that ${(i+1-n)}$ packets has been delivered during the $K$ time slots. \textcolor{black}{However, if the number of packets that could be delivered is greater than or equal to that is waiting to be transmitted, the minimum value of $n$ is $0$.} Hence, we can write the following recursive formulas for deriving $p_{n,G}$ and $p_{n,B}$ \textcolor{black}{in two cases}:
\begin{align}
\small
{p_{n,s}} = \left\{ {\begin{array}{*{20}{l}}
{\sum\limits_{{s^\prime } \in \{ G,B\} } {\sum\limits_{i = 0}^{K - 1} {{p_{i,{s^\prime }}}} } \sum\limits_{j = i + 1}^K {{P}} \left( {K,j,s|{s^\prime }} \right),}&{n = 0}\\
{\sum\limits_{{s^\prime } \in \{ G,B\} } {\sum\limits_{i = n - 1}^{K + n - 1} {{p_{i,{s^\prime }}}} } {P}\left( {K,i + 1 - n,s|{s^\prime }} \right),}&{n \ge 1}
\end{array}}, \right.\label{threeEle}
\end{align}

Theoretically, the values of $\{p_{n,G},n=0,1,\cdots\}$ and $\{p_{n,B},n=0,1,\cdots\}$ can be solved by combining \eqref{threeEle} with the factor
\begin{align}
\sum\limits_{n = 0}^\infty  {\left( {{p_{n,G}} + {p_{n,B}}} \right) = 1}. \label{summationEqu}
\end{align}
However, solving the above equations directly is rather challenging since the number of variables goes to infinity. 

\textcolor{black}{
\begin{example}
\emph{
We consider the special case where there is no memory, i.e., $p=r=0.5$. The value of $P(K,i,s^{\prime}|G)$ or $P(K,i,s^{\prime}|B)$ is equivalent to the probability that $(i-1)$ or $i$ slots being good state within $(K-1)$ slots, leaving the $(K+1)$th slot being state $s^\prime$, which could be expressed as
\begin{align}
&{P}\left( {K,i,G|G} \right) = {P}\left( {K,i,B|G} \right) = \left( {\begin{array}{*{20}{c}}
{K - 1}\\
{i - 1}
\end{array}} \right){2^{1 - K}}2^{-1},1 \le i \le K\\
&{P}\left( {K,i,G|B} \right) = {P}\left( {K,i,B|B} \right) = \left( {\begin{array}{*{20}{c}}
{K - 1}\\
i
\end{array}} \right){2^{1 - K}}2^{-1},0 \le i \le K - 1.
\end{align}
Then, we have
\begin{align}
&{p_{0,B}} = {p_{0,G}} = {2^{ - K}}\sum\limits_{i = 0}^{K - 1} {{p_{i,G}}\sum\limits_{j = i + 1}^K {\left( {\begin{array}{*{20}{c}}
{K - 1}\\
{j - 1}
\end{array}} \right)} }  + {2^{ - K}}\sum\limits_{i = 0}^{K - 1} {{p_{i,B}}\sum\limits_{j = i + 1}^K {\left( {\begin{array}{*{20}{c}}
{K - 1}\\
j
\end{array}} \right)} }  \label{eq:expression128}\\
&{p_{m,B}} = {p_{m,G}} = \sum\limits_{i = m - 1}^{K + m - 1} {\left[ {{p_{i,G}}\left( {\begin{array}{*{20}{c}}
{K - 1}\\
{i - m}
\end{array}} \right){2^{ - K}} + {p_{i,B}}\left( {\begin{array}{*{20}{c}}
{K - 1}\\
{i + 1 - m}
\end{array}} \right){2^{ - K}}} \right]} ,m \ge 1.\label{eq:129}
\end{align}
Denote by $p_n=p_{n,G}+p_{n,B}$. By symmetry, we have
\begin{align}
p_{n,G}=p_{n,B}=\frac{1}{2}p_{n},n=0,1,2,\cdots
\end{align}
Then, the equations in \eqref{eq:expression128} - \eqref{eq:129} can be simplified as
\begin{align}
&{p_0} = {2^{ - K}}\sum\limits_{i = 0}^{K - 1} {{p_i}\sum\limits_{j = i + 1}^K {\left( {\begin{array}{*{20}{c}}
K\\
j
\end{array}} \right)} }\\
&{p_m} = {2^{ - K}}\sum\limits_{i = m - 1}^{K + m - 1} {{p_i}\left( {\begin{array}{*{20}{c}}
K\\
{i + 1 - m}
\end{array}} \right)}
\end{align}
Consider the special case with $K=3$, we have
\begin{align}
\begin{array}{l}
{p_0} = 4{p_1}{\rm{ + }}{p_2}\\
8{p_1} = {p_0} + 3{p_1} + 3{p_2} + {p_3}\\
8{p_2} = {p_1} + 3{p_2} + 3{p_3} + {p_4}\\
8{p_m} = {p_{m - 1}} + 3{p_m} + 3{p_{m + 1}} + {p_{m + 2}}\\
 \cdots 
\end{array}\label{express1025}
\end{align}
The total number of equations in \eqref{express1025} is infinity since the AoI may grows unbounded. From the pattern in \eqref{express1025}, we can deduce that the sequence of variables $\{p_m,p_{m+1},...,\}$ for $m\geq 1$ forms a geometric sequence, i.e., we have
\begin{align}
p_{m+1}=\alpha p_m, m\geq 1,\label{express16}
\end{align}
then the equations in \eqref{express1025} for $m\geq1$ reduces to $1$ equation. With \eqref{express16} substituted into the $1$st and $3$rd equations of \eqref{express1025} and into \eqref{summationEqu}, we have
\begin{align}
\left\{ {\begin{array}{*{20}{l}}
{{p_0} = 4{p_1}{\rm{ + }}\alpha {p_1}}\\
{\left( {4 + \alpha } \right){p_1} + \frac{1}{{1 - \alpha }}{p_1} = 1}\\
{5 = \left( {4 + \alpha } \right) + 3\alpha  + {\alpha ^2}}
\end{array}} \right.
\end{align}
From which we can solve for the value of parameters as:
\begin{align}
\left\{ {\begin{array}{*{20}{l}}
{\alpha  = \sqrt 5  - 2}\\
{{p_0} = 3 - \sqrt 5 }\\
{{p_1} = 5\sqrt 5  - 11}
\end{array}} \right.
\end{align}
Next, we consider the general case of $K$, where we can solve for $p_0$, $p_1$ and $\alpha$ from the following equations
\begin{align}
&{{2^K}{p_0} = {p_0}\sum\limits_{j = 1}^K {\left( {\begin{array}{*{20}{c}}
K\\
j
\end{array}} \right)}  + {p_1}\left[ {\sum\limits_{j = 2}^K {\left( {\begin{array}{*{20}{c}}
K\\
j
\end{array}} \right)}  + \alpha \sum\limits_{j = 3}^K {\left( {\begin{array}{*{20}{c}}
K\\
j
\end{array}} \right)}  +  \cdots  + {\alpha ^{K - 2}}\sum\limits_{j = K}^K {\left( {\begin{array}{*{20}{c}}
K\\
K
\end{array}} \right)} } \right]}\notag\\
&{{p_0} + \sum\limits_{i = 0}^\infty  {{\alpha ^i}{p_1}}  = 1}\notag\\
&{{2^K}{p_1} = {p_0}\left( {\begin{array}{*{20}{c}}
K\\
0
\end{array}} \right) + {p_1}\left[ {\left( {\begin{array}{*{20}{c}}
K\\
1
\end{array}} \right) + \alpha \left( {\begin{array}{*{20}{c}}
K\\
2
\end{array}} \right) +  \cdots  + {\alpha ^{K - 1}}\left( {\begin{array}{*{20}{c}}
K\\
K
\end{array}} \right)} \right]},
\end{align}
which can be simplified to:
\begin{align}
&{p_0} = {p_1}\left[ {\sum\limits_{j = 2}^K {\left( {\begin{array}{*{20}{c}}
K\\
j
\end{array}} \right)}  + \alpha \sum\limits_{j = 3}^K {\left( {\begin{array}{*{20}{c}}
K\\
j
\end{array}} \right)}  +  \cdots  + {\alpha ^{K - 2}}\sum\limits_{j = K}^K {\left( {\begin{array}{*{20}{c}}
K\\
K
\end{array}} \right)} } \right]\notag\\
&{p_0} + \frac{1}{{1 - \alpha }}{p_1} = 1\notag\\
&{2^K}{p_1} = {p_0} + \frac{{{p_1}}}{\alpha }\left[ {{{\left( {1 + \alpha } \right)}^K} - 1} \right].
\end{align}
Combine the first and the third equation gives:
\begin{align}
\left( {K + 1} \right)\alpha  - {\alpha ^2}\sum\limits_{j = 3}^K {\left( {\begin{array}{*{20}{c}}
K\\
j
\end{array}} \right)}  -  \cdots  - {\alpha ^{K - 1}}\sum\limits_{j = K}^K {\left( {\begin{array}{*{20}{c}}
K\\
K
\end{array}} \right)}  = {\left( {1 + \alpha } \right)^K} - 1,
\end{align}
from which the parameter $\alpha$ can be solved numerically. Further, we get the results of $p_0$ and $p_1$.
\QEDA}
\end{example}}

\textcolor{black}{Based on the similar arguments presented in Example 1, we have the following relationship in general GE channel,}
\begin{align}
p_{n+1,s}=\beta p_{n,s}, \textnormal { for } n\geq 1, s\in\{G,B\}\label{alphaEqu},
\end{align}
where the factor $\beta\in(0,1)$ could be solved by substituting \eqref{alphaEqu} into \eqref{threeEle} with $n\ge2$. \textcolor{black}{With the factor about $p_{n,s}$ eliminated at both sides, we can solve for $\beta$ from the equation shown in Line~\ref{solveBeta} of Algorithm~\ref{alg1}.}

Further simplifying \eqref{threeEle} when $n=1$ renders
\begin{align}
p_{1,B} = \beta p_{0,B}\label{relation01},
\end{align}
which means that \textcolor{black}{\emph{the exponential trend of $p_{n,s}$ in \eqref{alphaEqu} beginning with $n=0$ for state $B$ and $n=1$ for state $G$.}}

Next, we only need to solve for \textcolor{black}{the initial probabilities, i.e.,}  $p_{0,G}$, $p_{1,G}$ and $p_{0,B}$ (Line~\ref{getThreeInt} in Algorithm~\ref{alg1}), which can be obtained  by the first two equations \textcolor{black}{($n=0$,$s=G$ or $s=B$)} in the recursive formula \eqref{threeEle}, together with \eqref{summationEqu}. Specifically,
\begin{align}
\mathbf{A}_{3\times3} = \left( {\begin{array}{*{20}{c}}
{1 - \sum\limits_{j = 1}^K {P\left( {K,j,G|G} \right)} },&{ - \sum\limits_{i = 0}^{K - 1} {{\beta ^i}\sum\limits_{j = i + 1}^K {P\left( {K,j,G|B} \right)} } },&{ - \sum\limits_{i = 0}^{K - 2} {{\beta ^i}\sum\limits_{j = i + 2}^K {{P_{GE}}\left( {K,j,G|G} \right)} } }\\
{\sum\limits_{j = 1}^K {P\left( {K,j,B|G} \right)} },&{\sum\limits_{i = 0}^{K - 1} {{\beta ^i}\sum\limits_{j = i + 1}^K {P\left( {K,j,B|B} \right)} }  - 1},&{\sum\limits_{i = 0}^{K - 2} {{\beta ^i}\sum\limits_{j = i + 2}^K {{P_{GE}}\left( {K,j,B|G} \right)} } }\\
1,&{\frac{1}{{1 - \beta }}},&{\frac{1}{{1 - \beta }}}
\end{array}} \right),
\end{align}
\begin{align}
\mathbf{b}={{{\left( {\begin{array}{*{20}{c}}
0&0&1
\end{array}} \right)}^T}}.
\end{align}

Given the values of $p_{0,G},p_{1,G}$, $p_{0,B}$ and $\beta$, we can obtain the general transmitter queue size and channel state distribution from \eqref{alphaEqu} and \eqref{relation01}. Substituting them back into \eqref{perE2Elatency} and then into \eqref{AoIperK}, we derive the average AoI with periodic arrival under the FCFS policy as
\begin{align}
\bar{\Delta}_{pF} ={p_{0,G}} +\frac{1}{{r\left( {1 - \beta } \right)}}\left( {1 + \frac{{r + \beta p}}{{1 - \beta }}} \right){p_{0,B}}+ \frac{1}{{1 - \beta }}\left( {1 + \frac{{p + r}}{{r\left( {1 - \beta } \right)}}} \right){p_{1,G}}+\frac{K-1}{2}  .\label{ETexpression}
\end{align}

\begin{algorithm}[htb]
\caption{average AoI numerical algorithm}
\label{alg1}
\begin{algorithmic}[1]
\STATE{\textbf{Input:}  $K,p,r$ };
\STATE{\textbf{Output:} $AoI$ };
\STATE{Calculate $P(K,i,s^{\prime}|s)$ recursively for $i=0,1,\cdots,K$ and $s,s^{\prime}\in\{G,B\}$};
\STATE{Solve $\beta$ with $\frac{{\beta  - \sum\limits_{i = 0}^K {{\beta ^i}P\left( {K,i,B|B} \right)} }}{{\sum\limits_{i = 0}^K {{\beta ^i}P\left( {K,i,G|B} \right)} }} = \frac{{\sum\limits_{i = 0}^K {{\beta ^i}P\left( {K,i,B|G} \right)} }}{{\beta  - \sum\limits_{i = 0}^K {{\beta ^i}P\left( {K,i,G|G} \right)} }}$, satisfying ${\beta\in(0,1)}$};\label{solveBeta}
\STATE{Solve $\mathbf{p}\triangleq {\left( {\begin{array}{*{20}{c}}
{{p_{0,G}}}&{{p_{0,B}}}&{{p_{1,G}}}
\end{array}} \right)^T}$ with $\mathbf{A}\mathbf{p}=\mathbf{b}$};\label{getThreeInt}
\STATE{Return the average AoI $\bar{\Delta}_{pF}$ with Eq~\eqref{ETexpression}}.
\end{algorithmic}
\end{algorithm}

\section{numerical results}\label{sec:results}
In this section, numerical results are provided to verify the analytical results for Bernoulli, periodic and generate-at-will arrival models under the  FCFS and pLGFS policies over the GE channel. The simulation results are obtained by averaging $10^3$ iterations and the network runs for $10^4$ time slots for each iteration.\footnote{The MATLAB codes for producing the results are available on Github at https://github.com/OUTMAN666/age-of-information-over-GE-channel.}

\begin{figure}[htb]
\begin{minipage}[t]{0.5\linewidth}
\centering
\includegraphics[width=8.2cm]{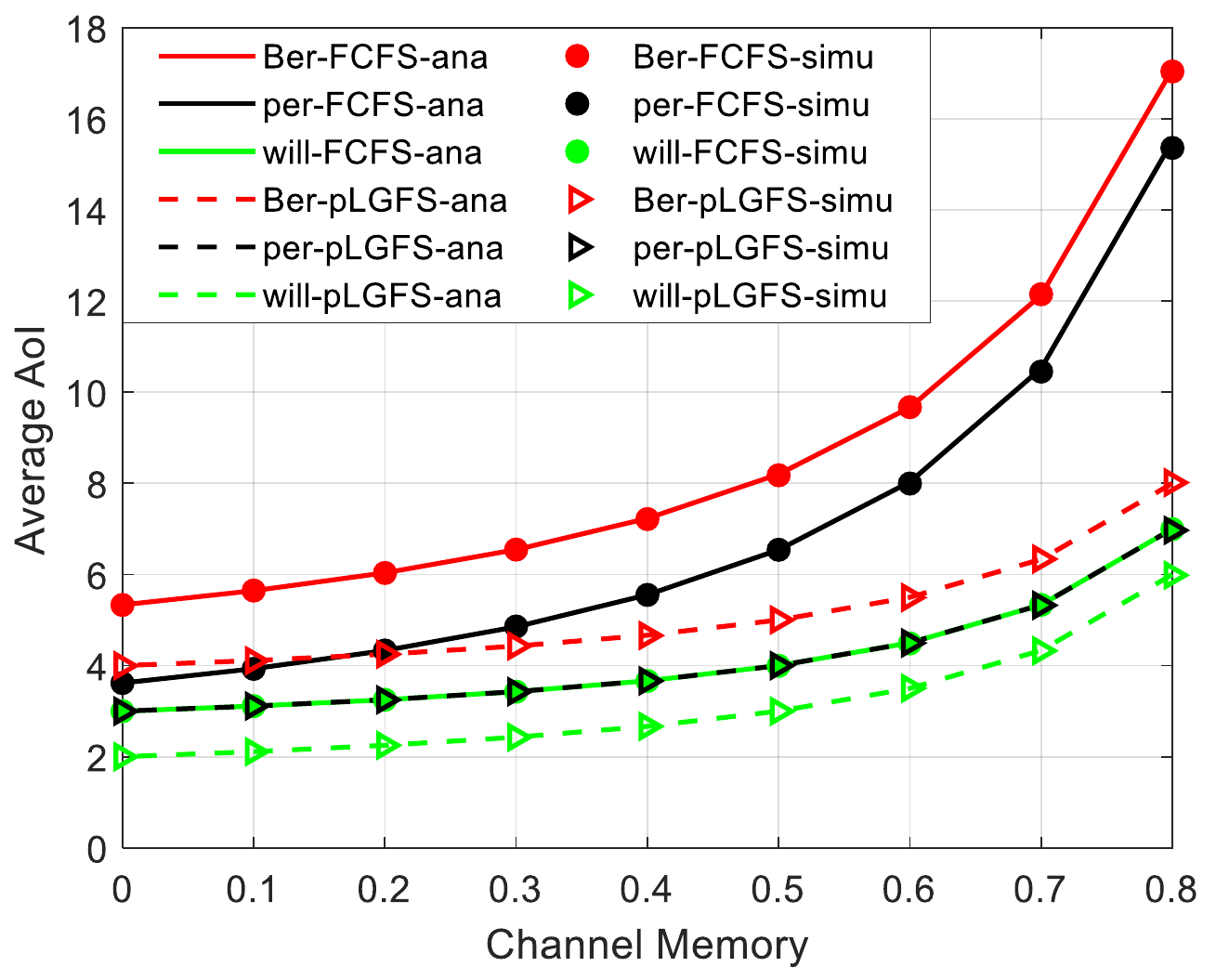}
\subcaption{$p_e^G=0$ and $p_e^B=1$ verification results.}
\label{fig:res01}
\end{minipage}%
\begin{minipage}[t]{0.5\linewidth}
\centering
\includegraphics[width=8.05cm]{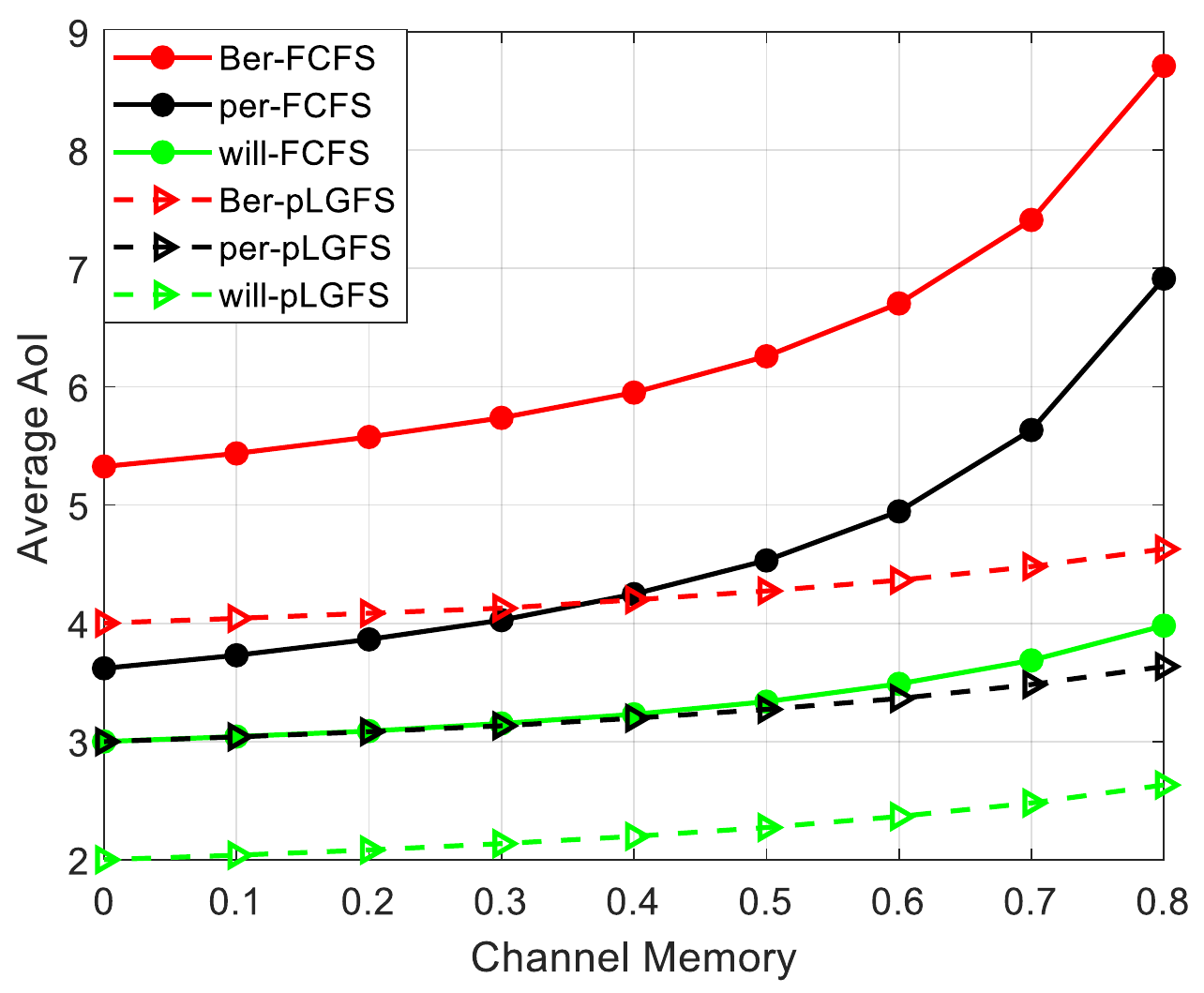}
\subcaption{$p_e^G=0.2$ and $p_e^B=0.8$ simulation results.}
\label{fig:res28}
\end{minipage}%
\caption{The impact of channel memory on the average AoI of various simulation results, with packet arrival rate $\lambda=\frac{1}{3}$ for Bernoulli arrival and $K=3$ for periodic arrival over the symmetric GE channel.}
\label{F:results}
\end{figure}

Fig.~\ref{fig:res01} compares the achievable average AoI under the FCFS and pLGFS policies, \textcolor{black}{with channel erasure probabilities $p_e^{G}=0$ and $p_e^{B}=1$.} The analytical AoI of the FCFS policy with Bernoulli and generate-at-will arrival models, the pLGFS policy with Bernoulli, periodic and generate-at-will arrival models under the symmetric GE channel are obtained by evaluating \eqref{eq:FCFSsys}, \eqref{eq:gen_at-will}, \eqref{eq:symLGFS}, \eqref{eq:pLGFSsys} and \eqref{eqx:generWillpLGFSsym}, respectively. It shows that the simulation results match very well with our analytical results in Section \ref{sec:Ber}, \ref{generate-at-will} and \ref{sec:Deter}. The pLGFS policy achieves lower AoI than the corresponding FCFS policy, at the cost of lower long-term average throughput. The average AoI of each scenario increases with channel memory monotonically. We deliberately set the same packet arrival rate, i.e., $\lambda=\frac{1}{K}=\frac{1}{3}$. \textcolor{black}{For the point-to-point channel considered in this letter,} periodic arrival model achieves lower AoI than Bernoulli arrival model, \textcolor{black}{which is different from that observed in the multi-user networks when the channel erasure probability is dominated by the interference among different users \cite{Emara}.} Further, the gap is $\frac{K-1}{2}$ for pLGFS according to the analysis. With memory increasing, the curve slope becomes steeper, and the average AoI under FCFS grows faster than that under pLGFS policy. \textcolor{black}{In Fig.~\ref{fig:res28}, we present the simulation results for GE channel with erasure probability $p_e^G=0.2$ and $p_e^B=0.8$. It is observed that the average AoI increases with the channel memory, with a smaller slope. This is because the difference between the good and bad states reduces when both the good and bad states experience non-zero erasure probabilities.}

\section{Conclusion}\label{sec:conclusion}
In this letter, we investigate the impact of channel memory on the data freshness. The closed-form expression of the average AoI under FCFS with Bernoulli and generate-at-will arrival models and that under pLGFS with Bernoulli, periodic and generate-at-will arrival models are derived. For FCFS with periodic arrival model, we propose an efficient numerical algorithm to calculate the average AoI. The analytical results reveal that the average AoI under pLGFS increases with channel memory $\eta$ at $\frac{\eta}{1-\eta}$ \textcolor{black}{over the symmetric GE channel}, while that under FCFS increases even faster with channel memory \textcolor{black}{due to the additional queueing delay at the transmitter}. The analysis approaches introduced in this letter could be applied for deriving the average AoI under other network settings, and the impact of channel memory could guide the design of more efficient AoI-optimal policy for networks with channel memory.

\begin{appendices}
\section{proof of PMF of system time}\label{appendixA}
Consider a typical segment of time, during which $k$ information packets entered the queue with probability $P_K(k)$. Denote $F_N(n)$ as the probability that the number of time slot being $n$, where $n \ge k $, we have
\begin{align}
{P_K(k)} = \sum\limits_{n = k}^\infty  {\left( {\begin{array}{*{20}{c}}
  n \\ 
  k 
\end{array}} \right){\lambda ^k}{{\left( {1 - \lambda } \right)}^{n - k}}F_N\left( n \right)} ,k = 0,1,2, \cdots
\end{align}
Denote $g(z)$ as the probability generating function (PGF) of $P_K(k)$, furtherly
\begin{align}
g\left( z \right) &= \sum\limits_{k = 0}^\infty  {\left( {\sum\limits_{n = k}^\infty  {\left( {\begin{array}{*{20}{c}}
  n \\ 
  k 
\end{array}} \right){\lambda ^k}{{\left( {1 - \lambda } \right)}^{n - k}}F\left( n \right)} } \right){z^k}}\notag\\
&= \sum\limits_{n = 0}^\infty  {{{\left( {\lambda z +  1 - \lambda } \right)}^n}F\left( n \right)}.
\end{align}
Let $y = \lambda z +1 - \lambda $,  then the PGF of $F_N(n)$ could be written as $\phi  \left( y \right) = \sum\limits_{n = 0}^\infty  {F_N\left( n \right){y^n}}$. We have
\begin{align}
g\left( z \right) = \phi \left( y \right) = \phi \left( {\lambda z + 1 - \lambda } \right)\label{eq:bridge}
\end{align}

For a system with FCFS queue discipline, the number of packets that the $n$th packet see left behind in the queue when it departs will be the number of arrivals that occurred while it was in the system \cite{bose2013introduction}, so does the PGF.

In the prevoius work \cite{Guan2112:Impact}, we have got the steady state probability distribution of  $G_n$ and $B_n$, corresponding to $n$ packets in the queue when a packet arrives at the transmitter with channel state being good and bad state respectively, where $n=0,1,2, \cdots$ We derive the PGF of the number of packets in the queue by combining the states two by two whose queue length dimensions are the same, i.e.,
\begin{align}
g\left( z \right) ={{G_0} - C + \frac{{{B_0} + C}}{{1 - {\frac{{p\lambda  + \left( {1 - p - r} \right)\lambda \left( {1 - \lambda } \right)}}{{\left( {1 - \lambda } \right)\left( {r + \left( {1 - p - r} \right)\lambda } \right)}}} z}}},\label{eq:gz}
\end{align}
\emph {where}
\begin{align}
{G_0} &= \frac{r}{{p + r}}\left( {1 - \frac{{p\lambda }}{{r\left( {1 - \lambda } \right)}}} \right),\\
{B_0} &= \frac{p}{{p + r}}\left( {\frac{{r - \left( {p + r} \right)\lambda }}{{\left( {1 - \lambda } \right)\left( {r + \left( {1 - p - r} \right)\lambda } \right)}}} \right),\\
C &\triangleq \frac{{p {G_0}}}{{p  + \left( {1 - p - r} \right) \left( {1 - \lambda } \right)}}.\label{Cdef}
\end{align}
Substitute \eqref{eq:gz} - \eqref{Cdef} and $z = \frac{{y + \lambda  - 1}}{\lambda }$ in \eqref{eq:bridge}, we have
\begin{align}
\phi \left( y \right) = {G_0} - C + \frac{{\left( {{B_0} + C} \right)\left( {r + \left( {1 - p - r} \right)\lambda } \right)}}{{1 - \frac{{p + \left( {1 - p - r} \right)\left( {1 - \lambda } \right)}}{{1 - \lambda }}y}}.
\end{align}
Together with the relationship that
\begin{align}
P_T(n) = F_N(n-1),
\end{align}
we derive the final result in \eqref{eq:Tn}.
\end{appendices}

\bibliographystyle{ieeetr}
\bibliography{AoIGE}

\end{document}